\documentclass[reprint,amsmath,amssymb, aps, superscriptaddress]{revtex4-2}
\pdfoutput=1
\usepackage[utf8]{inputenc}
\usepackage{graphicx}
\usepackage{float}
\usepackage{lipsum}  
\usepackage{xcolor}
\usepackage{braket}
\usepackage{physics}
\usepackage{xr}
\usepackage[colorlinks]{hyperref}

\usepackage[title]{appendix}

\begin{document}

\title{A Room-Temperature Extreme High Vacuum System for Trapped-Ion Quantum Information Processing}

\author{Lewis Hahn}
\email{rlhahn@uwaterloo.ca}
\thanks{These authors contributed equally to this work.}

\author{Nikhil Kotibhaskar}
\thanks{These authors contributed equally to this work.}

\author{Fabien Lefebvre}
\author{Sakshee Patil}
\author{Sainath Motlakunta}
\affiliation{Institute for Quantum Computing and Department of Physics and Astronomy, 
University of Waterloo, Waterloo, Ontario N2L 3G1, Canada}

\author{Mahmood Sabooni}
\affiliation{Institute for Quantum Computing and Department of Physics and Astronomy, 
University of Waterloo, Waterloo, Ontario N2L 3G1, Canada}
\affiliation{Open Quantum Design, Waterloo, Ontario N2L 6C2, Canada}

\author{Rajibul Islam}
\affiliation{Institute for Quantum Computing and Department of Physics and Astronomy, 
University of Waterloo, Waterloo, Ontario N2L 3G1, Canada}


\newcommand{\yb}{$^{171}\rm{Yb}^+\;$}
\newcommand{\ybtwo}{$^{172}\rm{Yb}^+\;$}
\newcommand{\q}{$\mathrm{mbar\ l\ s^{-1}\ cm^{-2}}$}
\newcommand{\Q}{$\mathrm{mbar\ l\ s^{-1}$}}

\begin{abstract}
We present a room-temperature Extreme High Vacuum (XHV) system engineered to support the long-duration operation of a trapped-ion quantum processor. Background-gas collisions impose limitations on trapped-ion performance and scalability by interrupting algorithmic execution and, in some cases, ejecting ions from the trap.
Using molecular-flow simulations, we optimize the chamber geometry, conductance pathways, and pumping configuration to maximize the effective pumping speed at the ion location.
We perform high-temperature heat treatment of stainless steel vacuum components to achieve the desired outgassing rate, guided by quantitative relations of bulk diffusive processes, allowing us to reduce the \(\mathrm{H_2}\) outgassing load to the $10^{-15}$ \q  level.
The final pressure in our chamber, measured by a hot cathode gauge, is $1.5\times10^{-12}\ \mathrm{mbar}$, corresponding to the gauge's measurement limit.
We measure the local pressure at the ion location by observing collision-induced reordering events in a long ion chain of mixed-isotope Yb\(^+\).
From the observed reordering frequency, we extract the average interval between collisions to be $(1.9 \pm 0.1)\ \mathrm{hrs/ion}$.
This corresponds to a local pressure of $(3.9 \pm 0.3)\times10^{-12}\ \mathrm{mbar}$ at the ion location, assuming that all collisions arise from background H\(_2\) molecules at room temperature.
Our demonstration extends the continuous operation time of a quantum processor while maintaining the simplicity of a room-temperature system that does not require cryogenic apparatus.
\end{abstract}
\maketitle 

\section{Introduction}
\label{sec:Introduction}
Trapped ions are among the most advanced and well-characterized platforms for quantum information processing (QIP), offering long coherence times, high-fidelity entangling gates, and programmable long-range interactions \cite{Wineland1998Review,Wang2021,Blatt2008Review,Haeffner2008}.
These ions are typically confined within an ultra-high vacuum (UHV) system to minimize collisions with background gases. 
As the number of trapped ions increases, the probability of such collisions also rises. 
Although quantum operations in current processors are not directly limited by these collisions, they still increase overhead to processor functionality. 
Even when high-energy collisions do not directly eject an ion, they can displace ions non-perturbatively from their equilibrium positions. 
In the presence of radio frequency (rf) fields used for trapping, the displaced ions can gain sufficient energy to escape the trap. 
Having to pause experiments to re-load ions can lead to changes in the operating conditions of the processor.
\begin{figure*}[hbt!]
  \centering
  \includegraphics[width=\textwidth]{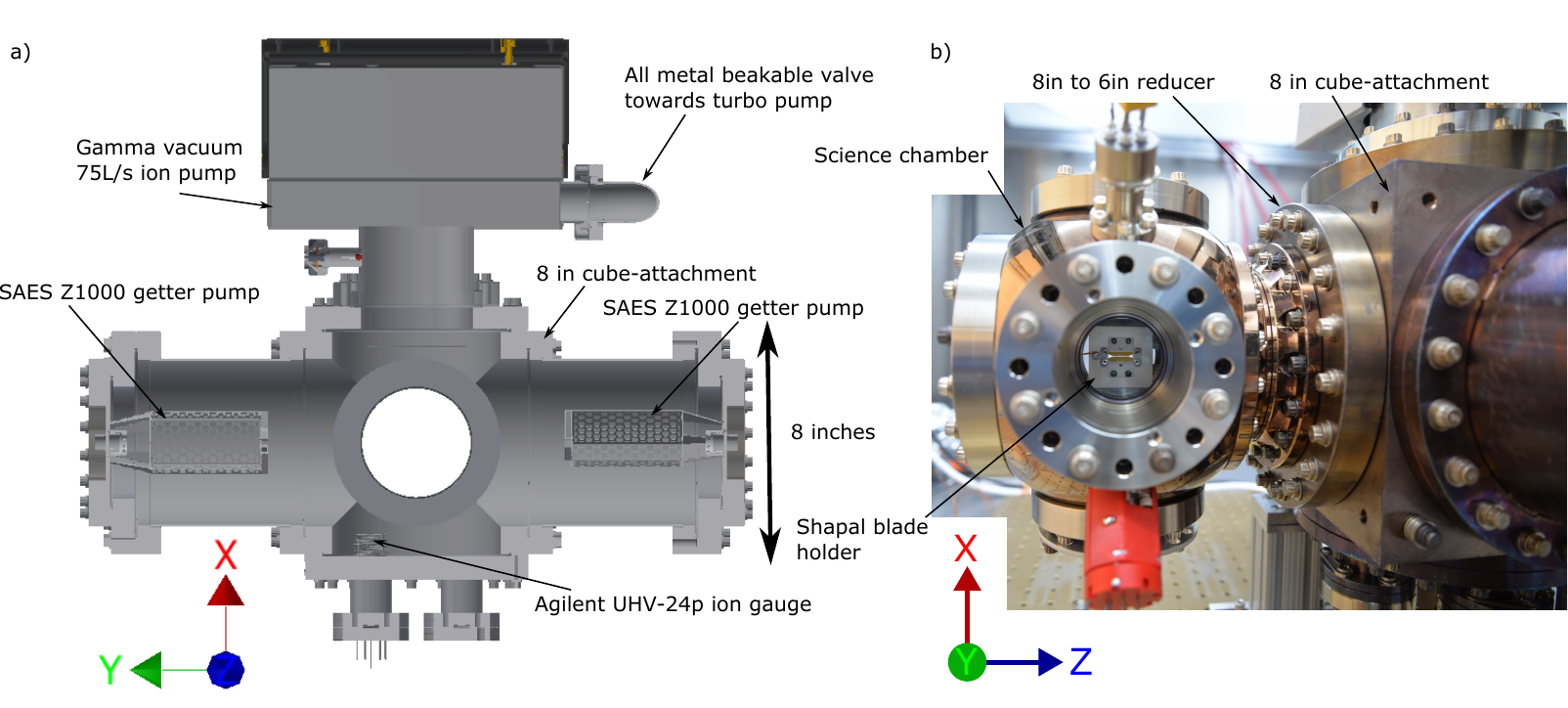}
  \caption{\textbf{Our vacuum system for long ion-chain experiments.} \textbf{Left:} CAD rendering of the 8~inch ConFlat cube-attachment showing the main pumping elements, including a 75~l/s Gamma ion pump, dual SAES~Z1000 non-evaporable getter (NEG) pumps, and an all-metal bake-able valve connected to a turbo-molecular pump for initial pump-down. 
  The Agilent UHV24p gauge was chosen for its low ($P<6.66\times10^{-12} \mathrm{mbar}$) x-ray limit specification. We found the x-ray limit to be below $1.5\times10^{-12} \mathrm{mbar}$. The coordinate axes indicate the ion trap's orientation within the science chamber (hidden behind the cube-attachment in this view). 
  \textbf{Right:} Photograph of the assembled chamber with the ion trap package mounted at the center of the science chamber, and its connection to the cube-attachment.
  The entire system has six optical viewports -- four re-entrant viewports (4.5 in diameter), and two other viewports (6 in diameter). 
  }
  \label{fig:pumping_internals}
\end{figure*}
For instance, the rf settings optimized for trapping might differ from those needed during quantum operations, forcing the system away from its steady operational state, potentially inducing drifts that require recalibration, and reducing the duty cycle.

A leading strategy to go beyond UHV, i.e., in the $10^{-12} \mathrm{mbar}$ regime or lower, is to house the vacuum chamber within a cryogenic environment \cite{Pagano2018ZigZag}. 
These systems typically operate around 4 K which leads to a reduction in collision frequency and energy. 
However, cryostat systems have major limitations and design challenges, such as decreased optical access, mechanical vibrations from the cooling systems, and cryogenic heat load from the internals of the vacuum system.
Here, we report on a room-temperature vacuum system at the boundary of extreme high vacuum (XHV), with the measured pressure of $1.5\times 10^{-12}\; \mathrm{mbar}$, limited by a hot-cathode gauge, and the local pressure at the ions to be $(3.9 \pm 0.3)\times 10^{-12} \;\mathrm{mbar}$ estimated from collisional reordering of an ion chain.
Our system (Fig. \ref{fig:pumping_internals}) enables four high-optical access windows for individual qubit manipulation, with long interval between collisions at $(1.9 \pm 0.1) \,\mathrm{hrs/ion}$ which is suited for quantum information processing with long-ion chains.

We rely on both numerical simulations, methodical preparation of in-vacuum components, including the adoption of best cleanroom practices, and quantitative heat treatment (see Sec. \ref{sec:Outgassing}) to realize the target low pressures.
The numerical molecular flow simulations allowed us to choose the system geometry, choice of vacuum pumps, and the target outgassing rates for all in-vacuum components.
Components of the vacuum system were then prepared following high-temperature heat treatment to hit the target outgassing rates.
The entire assembly was carried out in an ISO class 7 cleanroom following standard cleanroom practices.
This cleanroom also housed an industrial oven for baking the system without spatial temperature gradients to prevent unequal thermal expansion that may cause leaks.
All in-vacuum components were designed to prevent any virtual leaks as well.
We used two 1000 l/s non-evaporable getter (NEG) pumps in conjunction with a 75 l/s ion pump to achieve these pressures. 
We experimentally found that another commonly used pump -- Ti:sublimation pump -- failed to produce XHV, likely due to the presence of impurities in its Titanium filament \cite{Edwards1980MethanePump}.

In the standard conductance-limited regime, the pressure is given by,
\begin{equation}
    P = \frac{Q}{S_{\mathrm{eff}}},
    \label{eq:pressure_model}
\end{equation}
where $Q$ is the total outgassing load and $S_{\mathrm{eff}}$ is the effective pumping speed at the region of interest. 
In a well-prepared chamber, outgassing from internal surfaces dominates the overall gas load \cite{OHanlonVacuum}.
This motivates the introduction of the specific outgassing rate, $q = Q/A$, where $A$ is the surface area. 
In practice, the internal surface area $A$ is often limited by mechanical constraints such as optical access, the size and geometry of the ion trap etc. 
Consequently, we optimize the system to:
(i) reduce $q$ through rigorous material preparation, cleaning, and baking procedures, and  
(ii) maximize $S_{\mathrm{eff}}$ by minimizing conductance losses in vacuum pathways.

\section{Simulations}
\label{sec:Simulations}
While elaborate formulas exist for various geometries, real vacuum assemblies often include complex structures, distributed outgassing sources, and nonlocal pumping pathways. 
As a result, accurate prediction of system performance requires numerical simulation.
We employ Monte-Carlo–based molecular-flow simulations using MolFlow+ (CERN) \cite{Kersevan2019_Molflow_Synrad}, which tracks large ensembles of test particles undergoing diffuse reflections according to Knudsen’s cosine law \cite{knudsen1967cosine}. 
Pumping surfaces are represented by a specified sticking coefficient that directly corresponds to a pumping speed, and outgassing rates are applied to the internal surfaces. 
Because vacuum simulations are sensitive to input parameters, realistic outgassing rates and accurate geometric models are essential for extracting meaningful pressure values.
For an ion trap and many other atomic, molecular and optical (AMO) systems, the vacuum assembly can be segmented into three separate systems; the chamber which houses the experiment, the pumps and pump housing, and the attachments which are components used to connect the pumps and housings to the chamber. 
Separating the system in this way allows us to individually optimize each component without having to run a complex simulation with the full system. 

\begin{figure*}[hbt!]
  \centering
  \includegraphics[width=.8\textwidth]{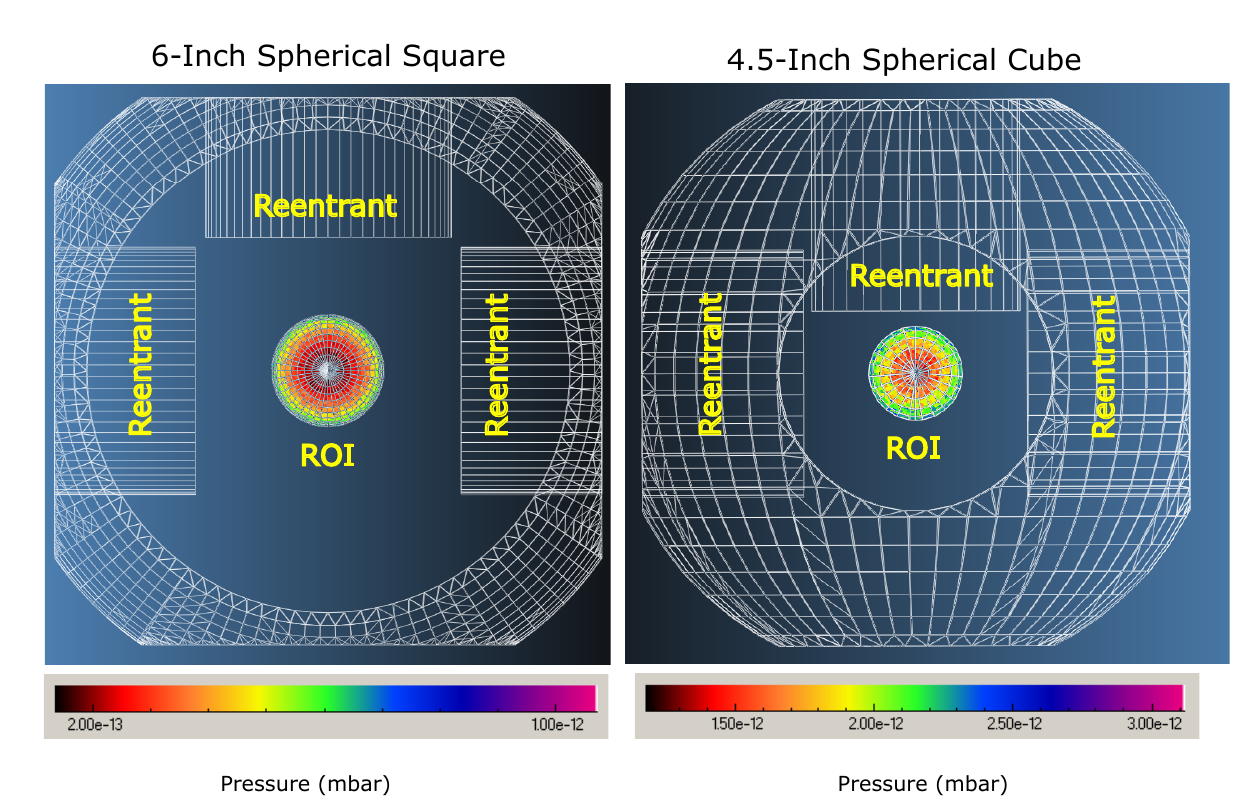}
  \caption{\textbf{Effect of vacuum chamber geometry on pressure.} Simulation results of the ultimate pressure in the Kimball physics 6-Inch spherical square (53-140320) and 4.5-Inch spherical cube (53-110400). 
  Here we have set a sticking factor = 1 on the pumping flange and a relatively high outgassing rate of $10^{-12}$ \q (expected from non heat-treated steel). 
  The average pressure at the ROI is 2.17$\times 10^{-12}$ vs 6.67$\times 10^{-13} \mathrm{mbar}$ respectively.}
  \label{fig:chamber}
\end{figure*}

While there may be options for chambers that meet the scientific requirements of the experiment, they can have significant effects on the final pressure of the system at the region of interest (ROI).
Figure~\ref{fig:chamber} shows two chambers that meet our scientific requirements of having four, $4.5 \ \mathrm{in}$ flanges for optical addressing.
To simulate the pressure at the ion location, we assume a sticking factor of 1 at the pumping flange, making the pumping speed equal to the conductance of the flange.
Because of the square dependence on orifice diameter (see Ref. \cite{NikhilPhD2024}), the conductance for the 6-Inch Spherical Square (left in Fig. \ref{fig:chamber}) is significantly higher than that of the 4.5-Inch Spherical Cube (right in Fig. \ref{fig:chamber}).
Assuming an outgassing rate of $10^{-12}$ \q, the best achievable pressure for the two chambers is $6.67\times 10^{-13}\ \mathrm{mbar}$ and $2.17\times 10^{-12}\ \mathrm{mbar}$ respectively, leading to our choice of the spherical square chamber (Fig. \ref{fig:pumping_internals}).
Our goal now is to design the rest of the system to maximize the pumping speed at the chamber's pumping flange.

\begin{figure}[h!]
  \centering
  \includegraphics[width=0.65\textwidth]{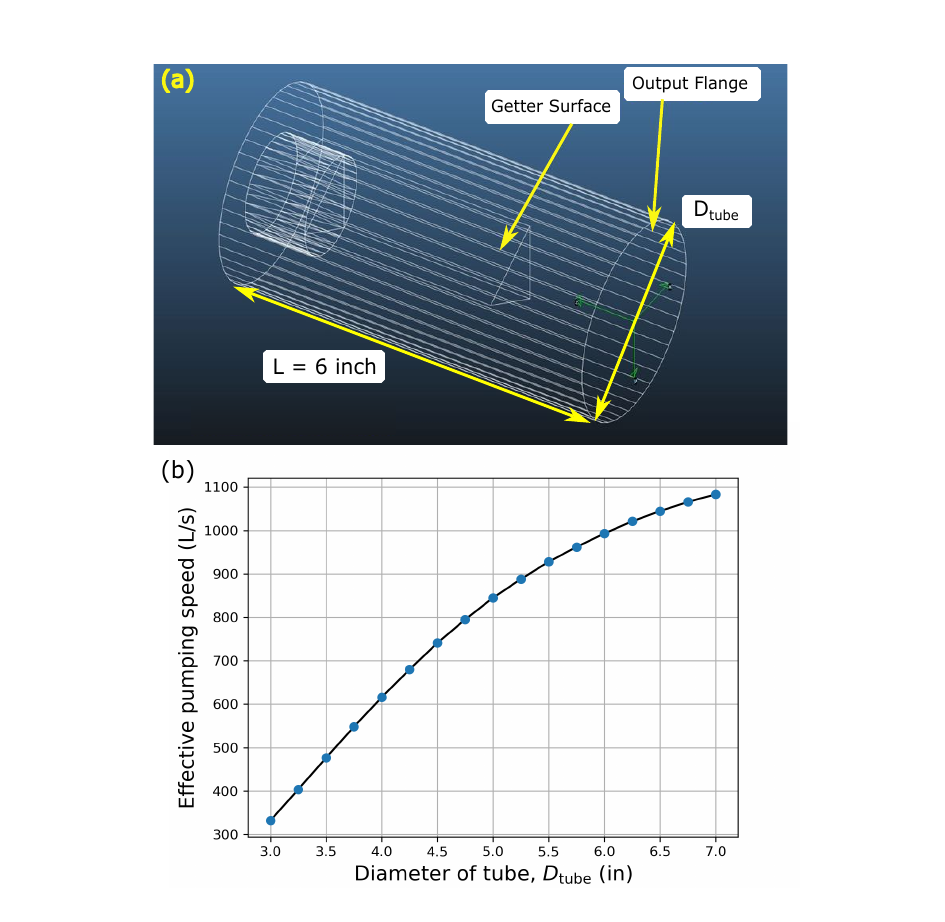}
  \caption{\textbf{The pumping speed of a pump depends on its housing diameter.} Simulation results from MolFlow+ of the effective pumping speed of a SAES Z1000 NEG pump inside a tube of fixed length ($\mathrm{L}=6\ \mathrm{in}$) and varying diameter, $D_\mathrm{tube}$. }
  \label{fig:pumping_speed}
\end{figure}

The most prominent pumps in an XHV system are non-evaporable getter (NEG) pumps. 
These come in a variety of sizes and pumping speeds reaching as high as $4000\ \mathrm{l/s}$. 
However, with poor conductance, the effective pumping speed at the ROI can be significantly reduced.
For our design, we chose the Z1000 NEG pump (from SAES Getters S.p.A) which has a pumping speed of $1250\ \mathrm{l/s}$ for hydrogen. 
We simulated a Z1000 inside tubes of varying diameter, $D_\mathrm{tube}$, and fixed length ($\mathrm{L}=6\ \mathrm{in}$) and measured the effective pumping speed at the opening of the tube. 
Figure~\ref{fig:pumping_speed} shows the results of this simulation.
From this, we see that a $D_\mathrm{tube} = 6\ \mathrm{in}$ tube has nearly a $3\times$ increase in effective pumping speed when compared to a $D_\mathrm{tube} = 3\ \mathrm{in}$ tube.
The length of the tube $L$ has a similar but less drastic affect on the effective pumping speed.

We also pick attachments connecting the pump housing to the pumping flange of the science chamber to give us the highest possible conductance.
We chose an $8$ in cube-attachment (Fig. \ref{fig:pumping_internals}) machined from a single piece of stainless steel (Highlight Tech Corp., Taiwan) over a 6-way cross as our simulations predict $\approx20 \%$ increase in pumping speed at the pumping flange. 

Once these optimizations are done, we simulate the entire system with all internal geometries. 
These simulations are vital for debugging when assembling the vacuum system and allow us to extract useful information like the difference in expected pressure at the gauge vs ion position.
Ultimately, for an outgassing rate of $10^{-14}$ \q, our simulations predict a pressure of $\approx5\times10^{-14}\ \mathrm{mbar}$ at the gauge location, and about three times higher at the ion location.

\section{Outgassing}
\label{sec:Outgassing}
With our vacuum components selected and the system simulated, we now have a target for the outgassing rate we must achieve to reach XHV pressures. 
We start by cleaning each component of the vacuum system by sonication in deionized water and detergent, followed by sonication in solvents such as acetone and isopropyl alcohol.
With a clean, leak-free system, there are two main sources of outgassing.
The first is surface outgassing, which consists primarily of water vapor which is desorped from the vacuum chamber walls. 
By performing a bake-out (typically around $200^\circ\mathrm{C}$) of the vacuum system, usually for between $2$ and $7$ days, the outgassing becomes dominated by the bulk diffusion of dissolved gases. 
This bulk outgassing consists primarily of hydrogen for stainless steel, copper and aluminum which are common metals used in UHV and XHV systems \cite{Chiggiato2020OutgassingAccelerators}.
During the manufacturing process, hydrogen is dissolved as single H atoms in molten metal and remains trapped in the bulk material when the metal solidifies. 
In a UHV system, these single H atoms diffuse to the surface of the metal and, after recombination, are released as molecular hydrogen. 

Heat treatment of vacuum chambers is a common practice in AMO communities for reaching UHV pressures. 
Typical procedures for treating stainless steel components include heating to $\approx 400^{\circ}\mathrm{C}$ under vacuum and/or in air \cite{Sefa2017,Calder1967Outgassing}.
These techniques result in $q_{\mathrm{H_2}} \approx 10^{-12}$ to $10^{-13}$ \q.
With room temperature ion trap systems having a surface area in the range of $100$ to $1000\ \mathrm{cm^2}$ and effective pumping speeds on the order of $10$ to $100\ \mathrm{l/s}$, typical pressures are in the $10^{-11}\ \mathrm{mbar}$ regime or higher \cite{Aikyo2020, Hankin2019}. 
To reach XHV pressures, we must decrease the outgassing rate to $q\approx10^{-14}$ to $10^{-15}$ \q. 
While this regime might seem very low, it is consistent with the measured outgassing rates in Ref. \cite{Sasaki2007Outgassing}.

For all stainless steel components, we then perform a $400^\circ\mathrm{C}$ heat treatment under vacuum using a pumping station consisting of a turbo and a roughing pump and heater tapes wrapped around the system for several days.
This heat treatment at high temperatures serves to reduce outgassing and also as the ultimate cleaning step, i.e., we do not perform any further chemical cleaning after the high-temperature heat treatment.
We give special considerations to the science chamber and the cube-attachment, as outgassing in the science chamber has a significant contribution to the local pressure at the ions, and the cube-attachment is the largest stainless steel component in the system.
For these, we adopt a quantitative approach by following known relations governing the outgassing of hydrogen from stainless steel.

We follow a diffusion-limited model \cite{Jousten2016HandbookVacuum,Jousten1961ThermalOutgassing} to estimate the outgassing rate of hydrogen from the vacuum walls at temperature $T$, given by:
\begin{equation}
\begin{aligned}\label{eqn:eqn2}
q(t) &\approx \frac{4D_{\mathrm{H_2}}(T)}{L}\times c_0 \sum^\infty_{n=0} \exp \left[-(2n+1)^2 \pi^2 \frac{D_{\mathrm{H_2}}(T)t}{L^2} \right] \\
&= \frac{4D_{\mathrm{H_2}}(T)}{L} c_0(t).
\end{aligned}
\end{equation}

Here, $q(t)$ is the specific outgassing as a function of time, $L$ is the thickness of the material, $c_0$ is the concentration of dissolved hydrogen measured in $\mathrm{mbar\ l\ cm^{-3}}$, 
\begin{equation}
\begin{aligned}\label{eqn:eqn3}
D_{\mathrm{H_2}}(T) = 4.7\times 10^{-3} e^{-\frac{0.56\mathrm{eV}}{k_B T}} 
\end{aligned}
\end{equation}
is the temperature-dependent diffusion coefficient of hydrogen in stainless steel, and $k_B$ is the Boltzmann's constant.
At room temperature (RT = 25$^\circ \mathrm{C}$), $D_{\mathrm{H_2}}\approx1\times10^{-12}\ \mathrm{cm^2\ s^{-1}}$.
The sum in Eqn. \ref{eqn:eqn2} determines how fast the concentration of dissolved gas decays over time. 
Instead of calculating the number of days we need to heat-treat our stainless steel components at high temperature, we instead proceed backwards from our target pressure at room temperature to estimate the pressure that we should observe at the end of the heat-treatment period while still at high temperature.

Equation \ref{eqn:eqn2} predicts that the concentration changes with time with a decay constant of many years at room temperature, and 26 days at high temperature (HT = 400$^\circ \mathrm{C}$), for $n=0$ and $L=25\ \mathrm{cm}$ -- both much longer than a day, which is the time it takes to ramp down from HT to RT.
Therefore, we can assume that the concentration of dissolved $\mathrm{H}_2$ remains constant during the temperature ramp down, and the ratio of the outgassing rates reduces to, attaching the suffixes HT and RT to variables defined earlier,
\begin{equation}
\begin{aligned}\label{eqn:eqn4}
\frac{q_{\mathrm{RT}}}{q_{\mathrm{HT}}} = \frac{D_{\mathrm{H}_2}(T_{\mathrm{RT}})}{D_{\mathrm{H}_2}(T_{\mathrm{HT}})},
\end{aligned}
\end{equation}
yielding $q_{\mathrm{RT}}\approx5\times 10^{-6} q_{\mathrm{HT}}$.
By combining Eqns. \ref{eq:pressure_model} and \ref{eqn:eqn4}, we can get an estimate for the pressure at HT immediately before the ramp down

\begin{equation}
\begin{aligned}\label{eqn:eqn5}
P_{\mathrm{HT}} = \frac{q_{\mathrm{RT}} A}{S_{\mathrm{eff}}} \times \frac{D_{\mathrm{H}_2}(T_{\mathrm{HT}})}{D_{\mathrm{H}_2}(T_{\mathrm{RT}})}.
\end{aligned}
\end{equation}
Given the surface area of our system and the pumping speed of the turbo pump used for heat treatment, the pressure during HT should reach \(5\times10^{-8}\ \mathrm{mbar}\) to achieve our target outgassing rate at room-temperature.
This leads us to adopt the strategy of either stopping the heat treatment once this pressure is reached or terminating it after a fixed number of days.

While the system was still at high temperature, we performed an air bake for 24 hours by turning off the pumping stations and venting atmospheric air before ramping down the temperature.
These air bakes have been shown to decrease the specific outgassing by a factor of two by forming an oxide layer on the surface of the stainless steel that acts like a diffusion barrier for hydrogen \cite{Westerberg1997Hydrogen,Sefa2017}.

In the end, both the science chamber and the cube-attachment were heat-treated, covering flanges with standard stainless steel blanks, for approximately 30 days, and the pressure reached $1.1\times10^{-8}$ and $4.8\times10^{-8}\ \mathrm{mbar}$ respectively.
These pressure values were extracted from an independent gauge connected to the vacuum system away from the high temperature region and using calculated values for the conductance between the gauge and the chamber location.
Using Eqn. \ref{eqn:eqn5} and accounting for the factor of two from the oxide layer we estimate the specific outgassings (at room temperature) of the science chamber and the cube-attachment to be,

\begin{center}
\begin{math}
\begin{aligned}
q_{\mathrm{chamber}} &< 4.58\times10^{-15}\ \mathrm{mbar\ l\ s^{-1}\ cm^{-2}}, \\
q_{\mathrm{cube}} &< 8.34\times10^{-15}\ \mathrm{mbar\ l\ s^{-1}\ cm^{-2}},
\end{aligned}
\end{math}
\end{center}

respectively, compatible with our target of reaching XHV in the final system.

However, a major bottleneck was outgassing from the mechanical holder of the ion trap electrodes, made from a machinable ceramic (Shapal from Precision Ceramics USA), for which the outgassing data was limited. 
Typically, dissolved water in the bulk of the ceramic is a major source of contamination.
By following the trend from short-duration heat treatments at $200^\circ\mathrm{C}$ and $400^\circ\mathrm{C}$ with a test piece of similar dimensions, we decided to heat-treat the mechanical holder at $900^\circ\mathrm{C}$ for 14 days in a tube furnace (OTF-1200X by MTI Corp) at $\approx1\times10^{-4}\ \mathrm{mbar}$.
This resulted in a base pressure at the same level as the empty cube-attachment.
Other components such as Kapton wrapped wires, ion trap electrodes, and Rogers PCB's were degassed during the final bake-out of the full system which was performed at $180^\circ\mathrm{C}$ (limited by the vacuum windows) for approximately 7 days.
After ramping the temperature down to room temperature and turning the ion pump on, our pressure gauge reached its saturation around $1.5\times 10^{-12}\  \mathrm{mbar}$ \cite{NikhilPhD2024}.
Further, by turning off the ion pump, and observing the increase in the pressure and its saturation due to pumping effects of the gauge itself, we estimate that the partial pressure of non-getterable gases in the system is $<1\times 10^{-12}\ \mathrm{mbar}$, see Appendix I for details.

\section{Pressure estimate with ions}
\label{sec:Pressure}
\begin{figure*}
    \centering
    \includegraphics[width=\textwidth]{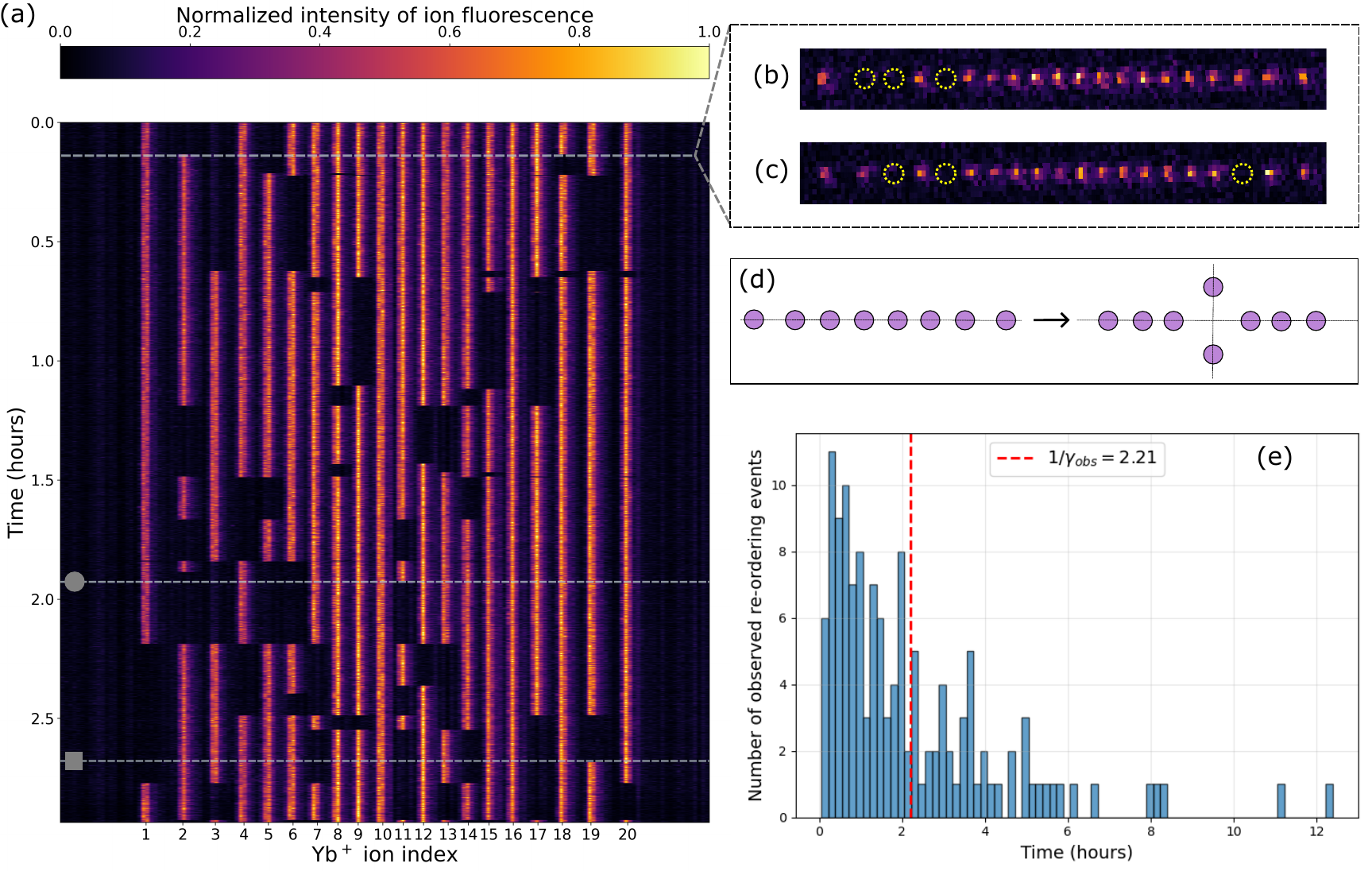}
    \caption{\textbf{Reordering of ions in a chain due to collisions.} \textbf{(a)} An example of a time-series heatmap of reordering events observed with a starting ion chain of 17 \yb ions and 3 dark ions, captured on a qCMOS camera every 5 seconds. The heatmap was generated by extracting a pixel-wide slice of the camera image, centered along the maximum intensity of the ion chain fluorescence. These slices were then stacked and displayed as the heatmap, with the y-axis being each frame's timestamp. Collision events are recorded by keeping track of changes in the line spanning each ion (corresponding to the ion index number). We count the instances when one of the \yb ions go dark due to possible Yb hydride formation (timestamped at $\bullet$) as a collision, but exclude the instances when one of the dark ions, in the presence of a high energy laser, returns to a bright state (timestamped at \scalebox{0.8}{$\blacksquare$}). \textbf{(b,c)} An example of the difference in the ion chain ordering, snapshot at the time indicated by the dashed line. The yellow dashed circles indicate the inferred positions of the dark ions, based on the spacing between ions. \textbf{(d)} A cartoon depiction of the two different ion chain configurations that we considered when estimating the minimum energy barrier $E_b$ for a reordering event. The left side shows the ions in their equilibrium axial position after minimizing the total energy $U$ of the chain (Appendix II Eqn. \ref{eqn:eqn9}); the right side shows the ion chain configuration we minimize the energy $U'$ for after constraining two ions to have the same axial position. \textbf{(e)} Histogram generated from extracting the time in between collision events, recorded from a heatmap such as the one in (a). The bin width used is 10 minutes, and the mean of the data is indicated by the red dashed line.}
    \label{fig:heatmap}
\end{figure*}

We estimate the pressure at the ions' location following the method outlined in Refs. \cite{Aikyo2020, Furst_2019}. 
We start by trapping a chain of Ytterbium ions, consisting primarily of \yb, with at least two being some other isotope which we refer to as dark ions. 
These are loaded independently by changing the frequency of the isotope-selective laser beam (399 nm).
Doppler cooling is applied resonantly for the \yb isotope, and the fluorescence captured on a qCMOS camera yields information about the locations of the \yb ions, while the locations of the dark ions can be inferred from the spacing between the bright ions.
We keep the Doppler laser-cooling beam on during the entire duration of the measurement.
We took five data sets and measured the time between 131 collision events to estimate the local pressure.
Each data set contains images captured at a regular interval, of either 5 or 10 seconds, for a total duration between 1 and 6 hours.
Figure \hyperref[fig:heatmap]{\ref*{fig:heatmap}a-c} visualizes a single data set showing the ion chain reordering from background collisions after melting and re-cooling.
The probability that a reordering of ions is distinguishable on the camera image is $p_{\mathrm{obs}} = 1 - \frac{n!(N-n)!}{N!}$ for $n$ dark ions in an $N$-ion crystal. 
For all of our datasets, $p\mathrm{_{obs}} > 0.99$.
We also must estimate the fraction of collisions that deliver enough energy for the ion chain to melt it in the first place, and use it (and $p\mathrm{_{obs}}$) to convert the observed reordering rate to the collision frequency.
For this, we assume that the collision is dominated by $\mathrm{H_2}$ molecules, and the energy imparted to an ion chain per elastic collision --  averaged over the Boltzmann distribution and the scattering angle -- is  $\langle \Delta E \rangle = 0.87~\mathrm{meV}$ \cite{Aikyo2020}.
The barrier energy for a reordering event is estimated by finding the minimum energy required to swap two ions through the radial plane and is shown pictorially in Fig. \hyperref[fig:heatmap]{\ref*{fig:heatmap}d}.

We estimate the trap frequencies to be $\omega_{x,y,z}/2\pi = (1.06,1.04,0.116) \ \mathrm{MHz}$. 
For a chain of 20 Ytterbium ions, the barrier energy for swapping is the highest at the edge of the chain, and estimated to be $\mathrm{E_b}  =0.35 \ \mathrm{meV}$ (see \hyperref[sec:Appendix_Barrier_energy]{Appendix II}).
As such, $p\mathrm{_{reorder}(E_b,\langle \Delta E \rangle)} = 88 \%$ of collisions will impart enough energy to cause a reordering (see Appendix II Eqn. \ref{eqn:p_reorder}).
Following Langevin theory \cite{Langevin1905}, pressure can be extracted from the collision rate $\gamma$ \cite{Aikyo2020},
\begin{equation}\label{eqn:eqn6}
  \gamma = \frac{\gamma_{\mathrm{obs}}}{p\mathrm{_{obs}}\cdot p\mathrm{_{reorder}}}=\frac{PQ}{k_BT}\sqrt{\frac{\pi\alpha_{\mathrm{H_2}}}{\mu\varepsilon_0}}.
\end{equation}
Here, $\gamma_{\mathrm{obs}}$ is the observed reordering rate, $P$ is the chamber pressure, $T=293 \ \mathrm{K}$ is the room temperature, $Q$ the net charge of the ion, $\alpha_{\mathrm{H_2}}=8.06 \times 10^{-31} \ \mathrm{m^3}$ \cite{Nie2025} is the polarizability of a hydrogen molecule, $\mu \approx m_{\mathrm{H_2}} = 3.35 \times 10^{-27} \ \mathrm{kg}$ the reduced mass and $\varepsilon_0$ is the vacuum permittivity.

From the gathered data, we extract a mean interval between collisions of $(1.9 \pm0.1)$ hrs/ion, where the observed reordering intervals are illustrated in Fig. \hyperref[fig:heatmap]{\ref*{fig:heatmap}e}.
Equation \ref{eqn:eqn6} yields a local ion pressure estimate of $(3.9\pm0.3)\times10^{-12}\ \mathrm{mbar}$ for the room-temperature vacuum system.
The error bar given for these two values corresponds to the standard error of the mean.

\section{Conclusions}
\label{sec:Conclusion}
In conclusion, we have reported a room-temperature ion trap apparatus operating at the boundary of extreme high vacuum, verified by both an independent pressure gauge and the trapped ions themselves.
These results represent roughly an order-of-magnitude improvement over typical room-temperature ion trap systems.
In this manuscript, we have presented our quantitative approach for selecting the system geometry and implementing outgassing mitigation techniques.
Although cryogenic systems can achieve lower pressures and collisions with background gas are, on average, less energetic and less detrimental, a room-temperature apparatus reduces overall design complexity, offers improved optical access for quantum control, and avoids vibrations associated with cryogenic infrastructure.
Our low collision intervals of \((1.9 \pm 0.1)\,\mathrm{hrs/ion}\) enable long operational times when working with extended ion chains, reducing downtime from retrapping events that may otherwise necessitate more frequent recalibration during complex quantum experiments.
Our approach can be applied to further optimize various parameters—such as chamber wall thickness and the geometry of connections between pumps and the science chamber—to achieve even lower pressures, up to fundamental physical limits.

\section*{Acknowledgments}

We thank Dr. Marcy Stutzman at the Thomas Jefferson Accelerator Facility for insights into XHV cleaning protocols, Ishwar Niraula from SAES for technical discussions, Shilpa Mahato for helping us on the experimental setup, and Mahmoud Badawy for help on the experimental control system and analysis.

We acknowledge financial support from the Canada First Research Excellence Fund (CFREF), the Natural Sciences and Engineering Research Council of Canada (NSERC) Discovery program (RGPIN-2018-05250 and RGPIN-2025-06496), the Government of Canada’s New Frontiers in Research Fund (NFRF), Ontario Early Researcher Award, University of Waterloo, and The Strategic Science Fund of the Government of Canada.

\section*{Author Contributions}

N.K. carried out numerical simulations and performed the system and protocol design for XHV operation. 
N.K. also led the heat treatment and baking of the apparatus, with support from S.M., L.H., and R.I..
N.K., S.M. and L.H. carried out the assembly of the vacuum system and internals.
L.H. led the ion-trapping efforts, performed the trapped-ion experiments, and developed the ion-based pressure measurement methods that validated the XHV performance. 
F.L., S.P., and M.S. assisted L.H. in building experimental subsystems and in data acquisition and analysis. 
R.I. supervised the entire project. 
All authors contributed to scientific discussions and to writing the manuscript.
Because N.K. and L.H. led the two main pillars of this work—designing the XHV system and validating it with trapped ions, respectively—they are recognized as co–first authors.

\section*{Appendix I: Pressure gauge and Non-getterable load}
\label{sec:Appendix_Pressure_gauge}
\begin{figure*}[htb!]
  \centering
  \includegraphics[width=\textwidth]{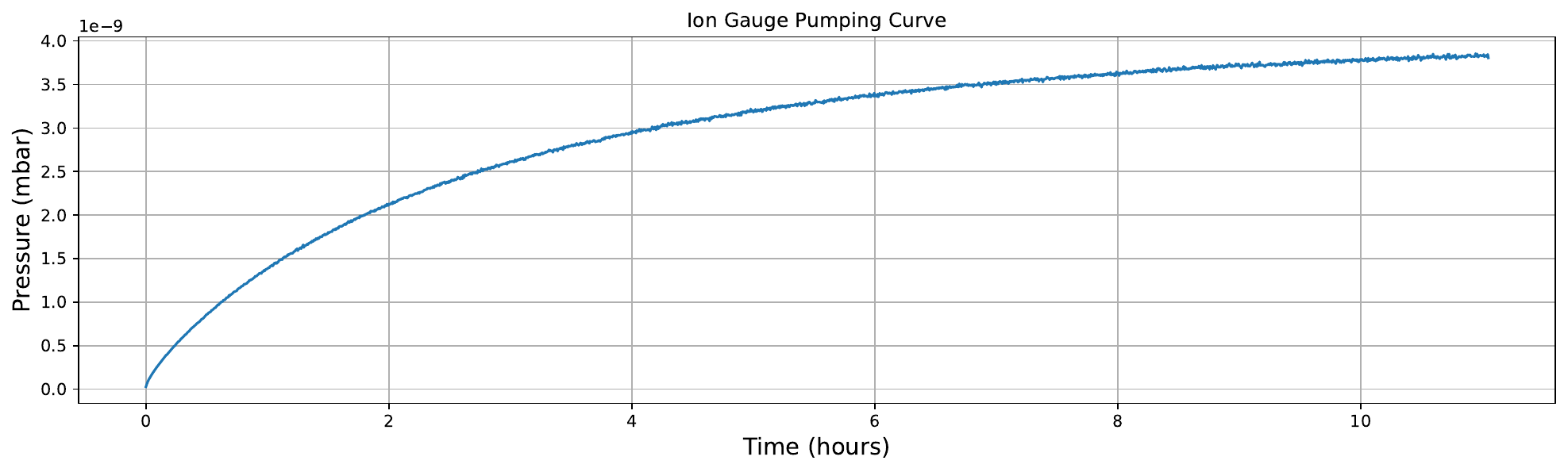}
  \caption{Pressure readings from the UHV24p gauge after turning off the ion pump. The pressure does not rise indefinitely due to the non-getterable pumping effects of the BA gauge.}
  \label{fig:ionpumpOFF}
\end{figure*}

For pressure monitoring, we use a hot cathode pressure gauge, also referred to as a Bayert Alpert (BA) gauge.
A BA gauge controller calculates pressure by ionizing gas molecules in a small volume and measuring the collector current, $I_c$, which is proportional to the number of ionized gas molecules. 
A current $I_e$ is passed through a filament, called an emitter, heating it. 
Electrons are then ejected from the filament and collide with gas molecules, ionizing them. 
The ionized molecules then interact with the collector, producing $I_c$, which depends on the background pressure $P$, emission current, and the gas-dependant sensitivity factor $S$ as:
\begin{equation}
\begin{aligned}\label{eqn:eqn7}
I_c = SPI_e,
\end{aligned}
\end{equation}
where we use the sensitivity factor for hydrogen as it is typically the dominant gas is UHV and XHV.

The ultimate pressure of the BA gauge depends on the lowest current readout of the controller and the x-ray limit of the gauge.
The UHV24p was chosen for its low x-ray limit, specified from the manufacturer to be $P<6.66\times10^{-12}\ \mathrm{mbar}$. 
We found the x-ray limit of our gauge to be even lower in practice, as we were consistently able to read pressures $P\approx1.5\times10^{-12}\ \mathrm{mbar}$.

While useful for monitoring pressure, BA gauges also allow us to extract other useful information about the system. 
Because a BA gauge ionizes and collects gas molecules, they have a pumping effect that depends on $I_e$.
This allows us to extract the non-getterable load in the system by turning off the ion pump and measuring the rate of pressure increase according to
\begin{equation}
\begin{aligned}\label{eqn:eqn8}
P(t) = P_{\mathrm{base}}+\frac{Q_{NG}}{S_g} \left(1-\exp\left( -\frac{S_g}{V}t \right) \right)
\end{aligned}
\end{equation}
Where $P_{\mathrm{base}}$ is the base pressure of the system, $S_g$ the pumping speed of the gauge, $Q_{NG}$ the outgassing rate of non-getterable molecules, and $V$ the volume of the system.

By fitting the data in Fig.~\ref{fig:ionpumpOFF} to Eqn.~\ref{eqn:eqn8}, we found $Q_{\mathrm{NG}}=6.15\times10^{-12}\,\mathrm{mbar\,L/s}$ and $S_g=1.6\times10^{-3}\,\mathrm{L/s}$ for an emission current of $I_e=4\,\mathrm{mA}$.

The Gamma vacuum ion pump in our system is expected to pump non-getterable gasses at a speed of $\approx10\ \mathrm{L/s}$.
This leads to an estimated partial pressure due to non-getterable load in our chamber to be $P_{NG}<1\times10^{-12}\ \mathrm{mbar}$. 
From our simulations and estimated outgassing, we expect the partial pressure due to getterable material to be lower and thus the limiting factor for the pressure will likely come from the non-getterable load. 

For a more detailed treatment of the content in this manuscript, including baking, prebaking, and cleaning procedures, see Ref \cite{NikhilPhD2024}.

\section*{Appendix II: Barrier energy}

\begin{figure}[h!]
  \centering
  \includegraphics[width=0.45\textwidth]{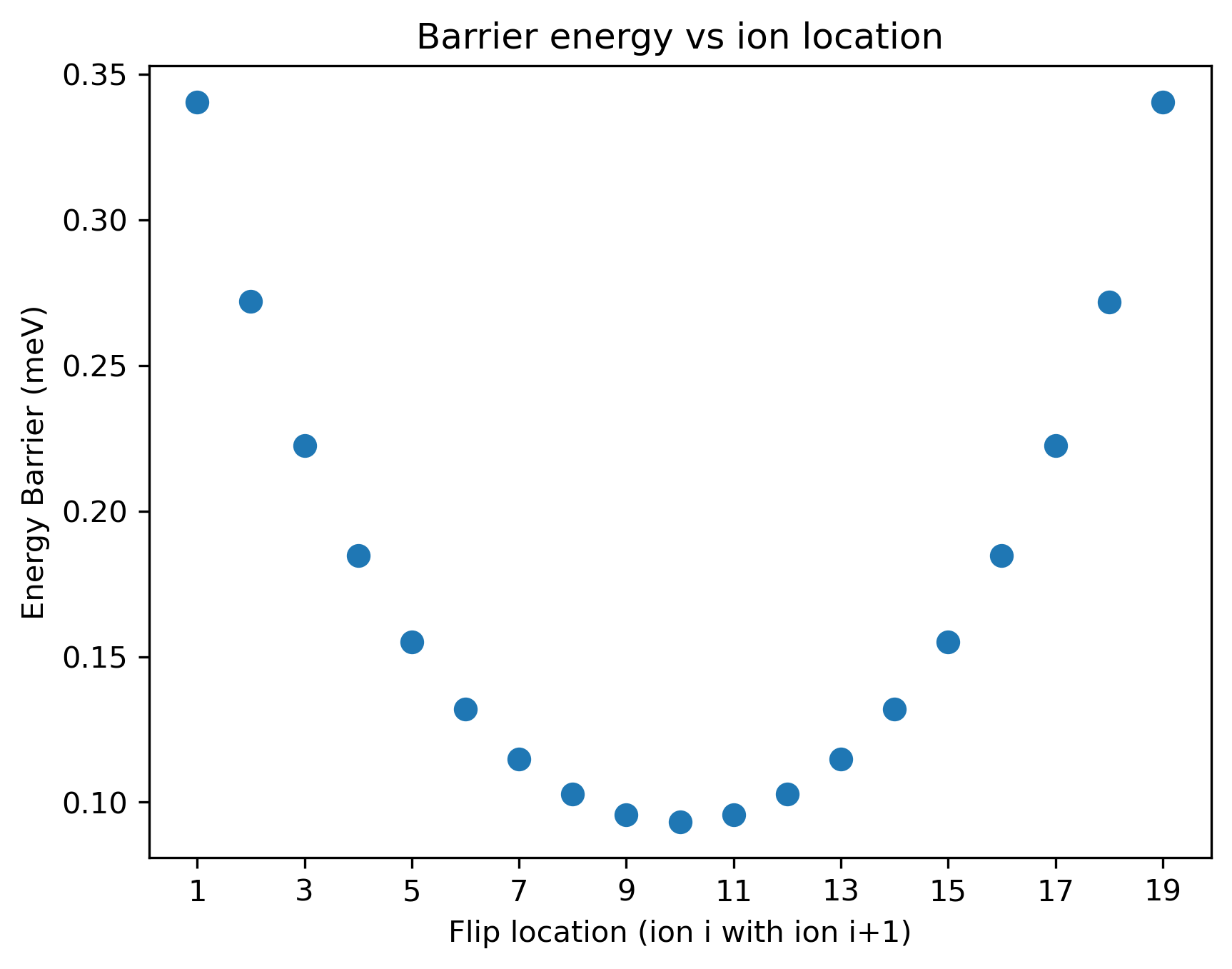}
  \caption{Barrier energy i(in $meV$) for flipping ions $i$ with $i+1$ in a 20 ion chain for trap frequencies $\omega_{x,y,z}/2\pi = (1.06,1.04,0.116) \mathrm{MHz}$.}
  \label{fig:Barrier_energy}
\end{figure}
\label{sec:Appendix_Barrier_energy}

We follow a similar structure to Ref. \cite{Aikyo2020} when calculating the berrier energy.
First, we find the equilibrium positions that minimize the total energy $U$ of the linear chain,
\begin{equation}
\begin{aligned}\label{eqn:eqn9}
U &= \sum_{i=1}^N\frac{m}{2}\left( \omega_x^2x_i^2+\omega_y^2y_i^2+\omega_z^2z_i^2\right) \\
 &+ \frac{Q^2}{4\pi\epsilon_0}\sum_{i=1}^{N-1}\sum_{j=i+1}^{N}\frac{1}{|{r_i-r_j}|},
\end{aligned}
\end{equation}
where $r_i =(x_i,y_i,z_i)$ are the positions of ion $i$, $\omega_{x,y,z}$ are the radial ($x,y$) and axial ($z$) trap frequencies, $Q$ and $m$ are the charge and mass of the ions respectively. 
Second, we constrain neighboring ions $i$ and ion $i+1$ to have the same axial positions, i.e. $z_i = z_{i+1}$ (see Fig. \hyperref[fig:heatmap]{\ref*{fig:heatmap}d}), and then optimize the spatial positions to minimize the energy $U'(z_i = z_{i+1})$.
We treat this as the energy requirement to swap neighboring ions $i$ and $i+1$.
We repeat this optimization for a swap of any two neighboring ions $i,i+1$ in the chain.
Figure \ref{fig:Barrier_energy} shows how the barrier energy changes for different ions flipping in the chain. 
We take the maximum $\max_{i}(U'(z_i = z_{i+1}) - U)$ as the barrier energy $E_b$ for a reordering event to occur.
The maximum barrier energy occurs at the outer most ions so $E_b=U'\left(z_1=z_2\right)-U$.
For a 20 ion chain and trap frequencies $\omega_{x,y,z}/2\pi = (1.06,1.04,0.116)\ \mathrm{MHz}$ we find $E_b\approx0.35\ \mathrm{meV}$.
Given that we are choosing a barrier energy that is large enough to swap any pair of ions, we are under-estimating the percentage of observed reordering events and thus over-estimating the collision frequency and hence the pressure. 

Following Ref. \cite{Spivey2021}, we find the probability that a background gas molecule at temperature $T$ has energy greater than $E_b$ as
\begin{equation}
\begin{aligned}\label{eqn:p_reorder}
p_{reorder}(E_b, T) = \frac{\int_{E_b}^\infty E \exp\left( \frac{-E}{k_B T} \right) d E}{\int_0^\infty E \exp\left( \frac{-E}{k_B T} \right) d E}
\end{aligned}
\end{equation}
The temperature $T$ is calculated using the average energy transfer given in the main text.

\newpage

\bibliographystyle{unsrt}
\bibliography{references}

@article{Wineland1998Review,
  title   = {Experimental issues in coherent quantum-state manipulation of trapped atomic ions},
  author  = {Wineland, D. J. and others},
  journal = {J. Res. Natl. Inst. Stand. Technol.},
  volume  = {103},
  number  = {3},
  year    = {1998},
  doi     = {10.6028/jres.103.019}
}

@article{Calder1967Outgassing,
  author  = {Calder, R. and Lewin, G.},
  title   = {Reduction of stainless-steel outgassing in ultra-high vacuum},
  journal = {British Journal of Applied Physics},
  volume  = {18},
  number  = {10},
  pages   = {1459},
  year    = {1967},
  month   = {Oct},
  doi     = {10.1088/0508-3443/18/10/313},
  url = {https://doi.org/10.1088/0508-3443/18/10/313},
}

@article{Blatt2008Review,
  title   = {Entangled states of trapped atomic ions},
  author  = {Blatt, R. and Wineland, D.},
  journal = {Nature},
  volume  = {453},
  pages   = {1008--1015},
  year    = {2008}, 
  doi     = {10.1038/nature07125},
  url     = {https://doi.org/10.1038/nature07125}
}

@article{Haeffner2008,
  author  = {H. Häffner and C.F. Roos and R. Blatt},
  title   = {Quantum computing with trapped ions},
  journal = {Physics Reports},
  volume  = {469},
  number  = {4},
  pages   = {155-203},
  year    = {2008},
  issn    = {0370-1573},
  doi     = {https://doi.org/10.1016/j.physrep.2008.09.003},
  url     = {https://www.sciencedirect.com/science/article/pii/S0370157308003463}
}

@book{OHanlonVacuum,
  title   = {A User's Guide to Vacuum Technology},
  author  = {O'Hanlon, J. F.},
  publisher = {Wiley},
  edition = {4},
  year    = {2003},
  doi     = {10.1002/0471467162}
}

@book{Jousten2016HandbookVacuum,
  editor    = {Jousten, K.},
  title     = {Handbook of Vacuum Technology},
  edition   = {2},
  publisher = {Wiley},
  year      = {2016},
  doi       = {10.1002/9783527688265},
  isbn      = {9783527413379},
  url       = {https://doi.org/10.1002/9783527688265}
}

@book{knudsen1967cosine,
  title={The Cosine Law in the Kinetic Theory of Gases},
  author={Knudsen, M.},
  series={NASA technical translation},
  url={https://books.google.ca/books?id=XEAnB-_MG-kC},
  year={1967},
  publisher={National Aeronautics and Space Administration}
}

@inproceedings{Jousten1961ThermalOutgassing,
  author    = {Jousten, K.},
  title     = {Thermal outgassing},
  booktitle = {Vacuum Technology: Proceedings of the CERN Accelerator School},
  series    = {CERN Yellow Reports: School Proceedings},
  number    = {CERN-99-05},
  publisher = {CERN},
  year      = {1999},
  pages     = {111--126},
  doi       = {10.5170/CERN-1999-005}
}

@article{Kersevan2019_Molflow_Synrad,
  author       = "Kersevan, Roberto and Ady, Marton",
  title        = "{Recent developments of Monte-Carlo codes Molflow+ and Synrad+}",
  reportNumber = "CERN-ACC-2019-173",
  pages        = "TUPMP037",
  year         = "2019",
  url          = "https://cds.cern.ch/record/2694236",
  doi          = "10.18429/JACoW-IPAC2019-TUPMP037",
}

@Article{Wang2021,
author={Wang, Pengfei and Luan, Chun-Yang and Qiao, Mu and Um, Mark and Zhang, Junhua and Wang, Ye and Yuan, Xiao and Gu, Mile and Zhang, Jingning and Kim, Kihwan},
title={Single ion qubit with estimated coherence time exceeding one hour},
journal={Nature Communications},
year={2021},
month={Jan},
day={11},
volume={12},
issn={2041-1723},
doi={10.1038/s41467-020-20330-w},
url={https://doi.org/10.1038/s41467-020-20330-w}
}

@article{Sefa2017,
  author  = {M. Sefa and J. Fedchak and J. Scherschligt},
  title   = {Investigations of medium-temperature heat treatments to achieve low outgassing rates in stainless steel ultrahigh vacuum chambers},
  journal = {Journal of Vacuum Science \& Technology A},
  volume  = {35},
  number  = {4},
  pages   = {041601},
  year    = {2017},
  url     = {https://tsapps.nist.gov/publication/get_pdf.cfm?pub_id=922647},
  doi     = {https://doi.org/10.1116/1.4983211}
}

@article{Aikyo2020,
  author  = {Y. Aikyo and G. Vrijsen and T. W. Noel and A. Kato and M. K. Ivory and J. Kim},
  title   = {Vacuum characterization of a compact room-temperature trapped ion system},
  journal = {Applied Physics Letters},
  volume  = {117},
  number  = {23},
  pages   = {234002},
  year    = {2020},
  issn = {0003-6951},
  doi     = {10.1063/5.0029236},
  url     = {https://doi.org/10.1063/5.0029236}
}

@phdthesis{Furst_2019,
  author       = {Henning Alexander Fürst},
  title        = {Trapped ions in a bath of ultracold atoms},
  school       = {Universiteit van Amsterdam},
  address      = {Amsterdam, Netherlands},
  year         = {2019},
  month        = {January},
  type         = {PhD Thesis},
  url          = {https://dare.uva.nl/search?identifier=b6294950-12e7-4b9f-a4b8-fd73df09707a}
}

@article{Spivey2021,
  title   = {High-Stability Cryogenic System for Quantum Computing With Compact Packaged Ion Traps},
  author    = {Spivey, Robert F. and Inlek, Ismail V. and Jia, Zhubing and Crain, Stephen and Sun, Ke and Kim, Junki and Vrijsen, Geert and Fang, Chao and Fitzgerald, Colin and Kross, Steffen and Noel, Thomas and Kim, Jungsang},
  journal = {IEEE Transactions on Quantum Engineering},
  year    = {2022},
  volume  = {3},
  pages   = {1--11},
  doi     = {10.1109/TQE.2021.3125926}
}

@article{Hankin2019,
  title = {Systematic uncertainty due to background-gas collisions in trapped-ion optical clocks},
  author = {Hankin, A. M. and Clements, E. R. and Huang, Y. and Brewer, S. M. and Chen, J.-S. and Chou, C. W. and Hume, D. B. and Leibrandt, D. R.},
  journal = {Phys. Rev. A},
  volume = {100},
  issue = {3},
  pages = {033419},
  numpages = {12},
  year = {2019},
  month = {Sep},
  publisher = {American Physical Society},
  doi = {10.1103/PhysRevA.100.033419},
  url = {https://link.aps.org/doi/10.1103/PhysRevA.100.033419}
}

@article{Pagano2018ZigZag,
  author        = {Pagano, G. and Hess, P. W. and Kaplan, H. and Tan, W. L. and Richerme, P. and Becker, P. and Kyprianidis, A. and Zhang, J. and Birckelbaw, E. and Hernandez, M.-L. and Wu, Y. and Monroe, C.},
  title         = {Cryogenic ion trap system for large scale quantum simulation},
  journal       = {Quantum Science and Technology},
  volume        = {4},
  number        = {1},
  pages         = {014004},
  year          = {2018},
  doi           = {10.1088/2058-9565/aae0fe}
}

@article{Langevin1905,
  author  = {P. Langevin},
  title   = {Une formule fondamentale de th{\'e}orie cin{\'e}tique},
  journal = {Ann. Chim. Phys.},
  volume  = {5},
  pages   = {245},
  year    = {1905}
}

@article{Nie2025,
  title = {Polarizability of molecular hydrogen and gas metrology},
  author = {Nie, Zhong-Liang and Wang, Jin and Hu, Chang-Le and Sun, Yu R. and Cheng, Cun-Feng and Tan, Yan and Hu, Shui-Ming},
  journal = {Phys. Rev. A},
  volume = {111},
  issue = {1},
  pages = {012801},
  numpages = {5},
  year = {2025},
  month = {Jan},
  publisher = {American Physical Society},
  doi = {10.1103/PhysRevA.111.012801},
  url = {https://link.aps.org/doi/10.1103/PhysRevA.111.012801}
}

@phdthesis{NikhilPhD2024,
  author = {N. Kotibhaskar},
  title  = {Towards Large Scale Quantum Simulations with Trapped Ions: Programmable XY model, Precise Light Sensing, and Extreme High Vacuum},
  school = {University of Waterloo},
  year   = {2024}
}

@article{Edwards1980MethanePump,
    title   = {Methane outgassing from a Ti sublimation pump},
    year    = {1980},
    journal = {Journal of Vacuum Science and Technology},
    author  = {Edwards, D.},
    number  = {1},
    month   = {1},
    pages   = {279--281},
    volume  = {17},
    publisher = {AIP Publishing},
    url     = {/avs/jvst/article/17/1/279/980054/Methane-outgassing-from-a-Ti-sublimation-pump},
    doi     = {10.1116/1.570412},
    url     = {https://doi.org/10.1116/1.570412},
    issn    = {0022-5355}
}

@article{Chiggiato2020OutgassingAccelerators,
    author  = {Chiggiato, Paolo},
    title   = {Outgassing properties of vacuum materials for particle accelerators},
    eprint  = {2006.07124},
    archivePrefix = {arXiv},
    primaryClass = {physics.acc-ph},
    month   = {6},
    year    = {2020},
    doi     = {10.48550/arXiv.2006.07124},
    url     = {https://doi.org/10.48550/arXiv.2006.07124}
}

@article{Westerberg1997Hydrogen,
  title   = {Hydrogen content and outgassing of air-baked and vacuum-fired stainless steel},
  author  = {Westerberg, L. and Hj{\"o}rvarsson, B. and Wall{\'e}n, E. and Mathewson, A.},
  journal = {Vacuum},
  volume  = {48},
  number  = {7},
  pages   = {771--773},
  year    = {1997},
  doi     = {https://doi.org/10.1016/S0042-207X(97)00042-0},
  url     = {https://www.sciencedirect.com/science/article/pii/S0042207X97000420}
}

@article{Sasaki2007Outgassing,
  author       = {Sasaki, Y. Tito},
  title        = {Reducing SS 304/316 hydrogen outgassing to $2\times10^{-15}$~Torr\,L\,cm$^{-2}$\,s$^{-1}$},
  journal      = {Journal of Vacuum Science \& Technology A},
  volume       = {25},
  number       = {4},
  pages        = {1309--1311},
  year         = {2007},
  month        = {07},
  issn         = {0734-2101},
  doi          = {10.1116/1.2734151},
  url          = {https://doi.org/10.1116/1.2734151}
}


\end{document}